\documentclass[aps,pre,reprint,superscriptaddress,showpacs,preprintnumbers,amsmath,amssymb,nofootinbib]{revtex4-1}
\usepackage[dvipdfmx]{graphicx}
\usepackage{dcolumn}   % needed for some tables
\usepackage{bm}        % for math
\usepackage{amssymb, amsmath, amsfonts}   % for math
\usepackage{lipsum}        % for math
\usepackage{color}
\DeclareGraphicsExtensions{{.pdf}}
\begin{document}
\newcommand{\noter}[1]{{\color{red}{#1}}}
\newcommand{\noteb}[1]{{\color{blue}{#1}}}
\newcommand{\field}{\left( \boldsymbol{r}\right)}
\newcommand{\paren}[1]{\left({#1}\right)}
\newcommand{\vect}[1]{\boldsymbol{#1}}
\newcommand{\uvect}[1]{\tilde{\boldsymbol{#1}}}
\newcommand{\vdot}[1]{\dot{\boldsymbol{#1}}}
\newcommand{\vder}{\boldsymbol{\nabla}}
%
% The following information is for internal review, please remove them for submission
\widetext
%\leftline{Primary authors: Norihiro Oyama}
%\leftline{To be submitted to Frontiers in Physics}
%
% the following line is for submission, including submission to the arXiv!!
%\hspace{5.2in} \mbox{Fermilab-Pub-04/xxx-E}
%
\title{
  Dynamic susceptibilities in dense soft athermal spheres under a finite-rate shear
}
%\input author_list.tex       % D0 authors (remove the first 3 lines
                             % of this file prior to submission, they
                             % contain a time stamp for the authorlist)
% (includes institutions and visitors)
\author{Norihiro Oyama}
\email{oyamanorihiro@g.ecc.u-tokyo.ac.jp}
\affiliation{Graduate School of Arts and Sciences, The University of Tokyo, Tokyo 153-8902, Japan}
\affiliation{Mathematics for Advanced Materials-OIL, AIST, Sendai 980-8577, Japan}
\author{Takeshi Kawasaki}
\email{kawasaki@r.phys.nagoya-u.ac.jp}
\affiliation{Department of Physics, Nagoya University, Nagoya 464-8602, Japan}
\author{Kuniyasu Saitoh}
\affiliation{Department of Physics, Faculty of Science, Kyoto Sangyo University, Motoyama, Kamigamo, Kita-ku, Kyoto 603-8555, Japan}

\date{\today}
\begin{abstract}
The mechanical responses of dense packings of soft athermal spheres under a finite-rate shear are studied by means of molecular dynamics simulations.
{We investigate the volume fraction and shear rate dependence of the fluctuations in the shear stress and the interparticle contact number.}
In particular, we quantify them by defining the susceptibility as the ratio of the global to local fluctuations. 
{The obtained susceptibilities form ridges on the volume fraction-shear rate plane, which are reminiscent of the Widom lines around the critical point in an equilibrium phase transition.}
\end{abstract}
\maketitle

%
%%%%%%%%%%%%%%%%%%%%%%%%%%%%%%%%%%%%%%%%%%%%%%%%%%%%%%%%%%%%%%%%%%%%%%%%%%%%%%%%%%%%%%%%%%%%%%%%%%%%%%%%%%%%%%%%%%%%%%%%%%%%%%
% Introduction
%%%%%%%%%%%%%%%%%%%%%%%%%%%%%%%%%%%%%%%%%%%%%%%%%%%%%%%%%%%%%%%%%%%%%%%%%%%%%%%%%%%%%%%%%%%%%%%%%%%%%%%%%%%%%%%%%%%%%%%%%%%%%%
\section{Introduction}
%%% 1. Soft athermal spheres and jamming (in general)
Soft condensed matters comprising bubbles, emulsions, or powder particles are generally referred to as ``soft athermal particle systems". Soft athermal particles are characterized by their (quasi-) elastic interactions, and thermal motion is negligible since they are large in size. When their density increases quasistatically, a transition from the liquid state, where the stress is zero, to the amorphous solid state, where the stress is finite, occurs. This transition is called the jamming transition~\cite{Andrea1998,OHern2003}. In the vicinity of the jamming transition point, various physical quantities, namely, the stress, the interparticle contact number, and the viscosity, behave critically~\cite{Durian1995,VanHecke2010,Kawasaki2015Phys.Rev.E,Olsson2019Phys.Rev.Lett.,Ikeda2020Phys.Rev.Lett.,Saitoh2020b}. The jamming transition is similar to the glass transition observed in thermal particle systems such as atomic, molecular, and colloidal systems; recently, however, they have been revealed to be distinct~\cite{,Ikeda2012,Ikeda2013}.

%%% 2. Brief review on the study on rheology 
The rheology of athermal particles with shear flow also exhibits critical behaviors caused by the jamming transition. In particular, a scaling function for the flow curve regarding the volume fraction and shear rate has been proposed~\cite{Olsson2007Phys.Rev.Lett.}, and the validity of the scaling has been widely discussed to date~\cite{Kawasaki2015Phys.Rev.E,Vagberg2016,Bonn2017Rev.Mod.Phys.,Saitoh2020}. First, the jamming transition can be strictly defined in the athermal quasistatic limit; thus, under a finite-rate shear, the existence of a jamming transition is not obvious. Most conventional jamming transition studies are concerned with the criticality of macroscopic mean quantities, whereas with the finite-rate shear, physical quantities such as the shear stress continuously increase with increasing volume fraction, and no remarkable singularity is observed~\cite{Heussinger2009Phys.Rev.Lett.,Vagberg2014,Vagberg2016,Vescovi2016a,Nagasawa2019}. In the statistical mechanics of thermal equilibrium systems, a naive phase transition picture is often captured by the fluctuation of physical quantities. In previous studies on the jamming transition, little discussion on the fluctuation has been made, although it is potentially significant. Accordingly, this work focuses on the fluctuation of the physical quantities and clarifies the jamming transition behavior under a finite shear rate.

%%% 3. In this work...
{In this work, we investigate the stress response of soft athermal particles using molecular dynamics simulations with a finite-rate shear flow. We measure the volume fraction dependence of the shear stress under a constant shear rate, and then, near the jamming transition point, which is characterized by the athermal quasistatic (AQS) limit, we find that the fluctuation of the stress exhibits a peak. We also find that the peak height diverges and the peak position converges to the jamming transition point when we decrease the shear rate towards the AQS limit, which is reminiscent of the Widom line near the critical point in an equilibrium phase transition. Despite this similarity, the mechanism of these fluctuations in dense athermal particles is still not apparent due to their strong nonequilibriumness. Hence, to clarify the mechanism, we investigate the time evolution of the stress when the stress fluctuation is enhanced, and we reveal that under a wide range of finite rates, the system transiently acquires rigidity intermittently. We furthermore obtain the Widom line from the contact number fluctuations, which converge to the jamming transition point in the AQS limit, yet its trace is not identical to that of the stress fluctuation. These findings deepen our understanding of the jamming transition under a finite-rate shear and provide us with extensible knowledge for various phase transition phenomena under an external field.

%%% 4. Construction of the paper...
This paper is constructed as follows. First, we introduce the numerical simulation method. Next, we discuss the average shear stress and its fluctuation. Then, we examine the stress-strain curve and contact number fluctuations. Afterward, we draw the Widom lines obtained from the stress and contact number fluctuations. Finally, we summarize the results and give our perspectives.}

%%%%%%%%%%%%%%%%%%%%%%%%%%%%%%%%%%%%%%%%%%%%%%%%%%%%%%%%%%%%%%%%%%%%%%%%%%%%%%%%%%%%%%%%%%%%%%%%%%%%%%%%%%%%%%%%%%%%%%%%%%%%%%
% Method
%%%%%%%%%%%%%%%%%%%%%%%%%%%%%%%%%%%%%%%%%%%%%%%%%%%%%%%%%%%%%%%%%%%%%%%%%%%%%%%%%%%%%%%%%%%%%%%%%%%%%%%%%%%%%%%%%%%%%%%%%%%%%%
\section{Numerical methods}
We employ molecular dynamics (MD) simulations of soft athermal particles in three dimensions.
To avoid crystallization of the system, we prepare a 50:50 binary mixture of $N$ particles,
where different kinds of particles have the same mass $m$ and different diameters, $d$ and $1.4 d$~\cite{OHern2003}.
The force between the particles, $i$ and $j$, in contact is modeled by a ``linear spring-dashpot"~\cite{Luding2005},\
i.e.,\ $\bm{f}_{ij}=(k\xi_{ij}-\eta \dot{\xi}_{ij})\bm{n}_{ij}$, with the stiffness $k$ and viscosity coefficient $\eta$.
The force is parallel to the normal unit vector $\bm{n}_{ij}=\bm{r}_{ij}/|\bm{r}_{ij}|$,
where $\bm{r}_{ij}\equiv\bm{r}_i-\bm{r}_j$, with the particle positions, $\bm{r}_i$ and $\bm{r}_j$ denoting the relative positions.
In addition, $\xi_{ij}=R_i+R_j-|\bm{r}_{ij}|>0$ is the overlap between the particles, and $\dot{\xi}_{ij}$ is its time derivative,
where $R_i$ ($R_j$) is the radius of particle $i$ ($j$).
The stiffness and viscosity coefficient determine the time scale as $t_0\equiv\eta/k$
and are adjusted such that the \emph{normal restitution coefficient} of the particles is exactly zero,\
i.e., $e=\text{exp}(-\pi/\sqrt{2mk/\eta^2-1})=0$~\cite{Luding2005}.

We randomly distribute the $N$ particles in an $L\times L\times L$ cubic periodic box and relax the system to a mechanically stable state~\cite{Bitzek2006}.
% Shear
Then, we apply simple shear deformations to the system under the Lees-Edwards boundary conditions~\cite{Lees1972}.
In each time step, we apply affine deformation to the system by replacing every particle position $(x_i,y_i,z_i)$ with $\bm{r}_i=(x_i+\Delta\gamma y_i,y_i,z_i)$ ($i=1,\dots,N$)
and then numerically integrate the equations of motion, $m\ddot{\bm{r}}_i=\sum_j\bm{f}_{ij}$, with a small time increment $\Delta t$~{\cite{Saitoh2016b,Saitoh2017}}.
Here, $\Delta\gamma$ is the strain increment; hence, the shear rate is defined as $\dot{\gamma}\equiv\Delta\gamma/\Delta t$.

% Data
In our MD simulations, we control the volume fraction of the particles $\varphi$ and the shear rate $\dot{\gamma}$.
To control the shear rate, we change both $\Delta\gamma$ and $\Delta t$ within the constraints\
$\Delta \gamma\le 10^{-6}$ and $\Delta t \le 0.1 t_0$.
In addition, we measure the mechanical responses of the system to simple shear deformations by the shear stress
\begin{equation}
    \sigma = -\frac{1}{L^3}\sum_{i,j}f_{ijx}^\mathrm{el}r_{ijy}~.
    \label{eq:sigma}
\end{equation}
Here, $f_{ijx}^\mathrm{el}=k\xi_{ij}n_{ijx}$ is the $x$-component of the elastic force,
and $r_{ijy}$ is the $y$-component of the relative position $\bm{r}_{ij}$ between the particles $i$ and $j$, which are in contact.
For each $\varphi$ and $\dot{\gamma}$, we compute the mean value $\langle\sigma\rangle$ and fluctuations of the shear stress in a steady state,
where the applied strain is in the range $1<\gamma<5$.
We also take ensemble averages of $\langle\sigma\rangle$ and $\chi_\sigma$ (the definitions of which are given in Sec.~\ref{sec:chi}) over at least $20$ different initial configurations.
%
%
%%%%%%%%%%%%%%%%%%%%%%%%%%%%%%%%%%%%%%%%%%%%%%%%%%%%%%%%%%%%%%%%%%%%%%%%%%%%%%%%%%%%%%%%%%%%%%%%%%%%%%%%%%%%%%%%%%%%%%%%%%%%%%
% Results
%%%%%%%%%%%%%%%%%%%%%%%%%%%%%%%%%%%%%%%%%%%%%%%%%%%%%%%%%%%%%%%%%%%%%%%%%%%%%%%%%%%%%%%%%%%%%%%%%%%%%%%%%%%%%%%%%%%%%%%%%%%%%%
\section{Results}
%------------------------------------------------------------------------------------------
%%% Fig. 1. Averages and fluctuations of stress as functions of the volume fraction
%------------------------------------------------------------------------------------------
\begin{figure}
\includegraphics[width=\linewidth]{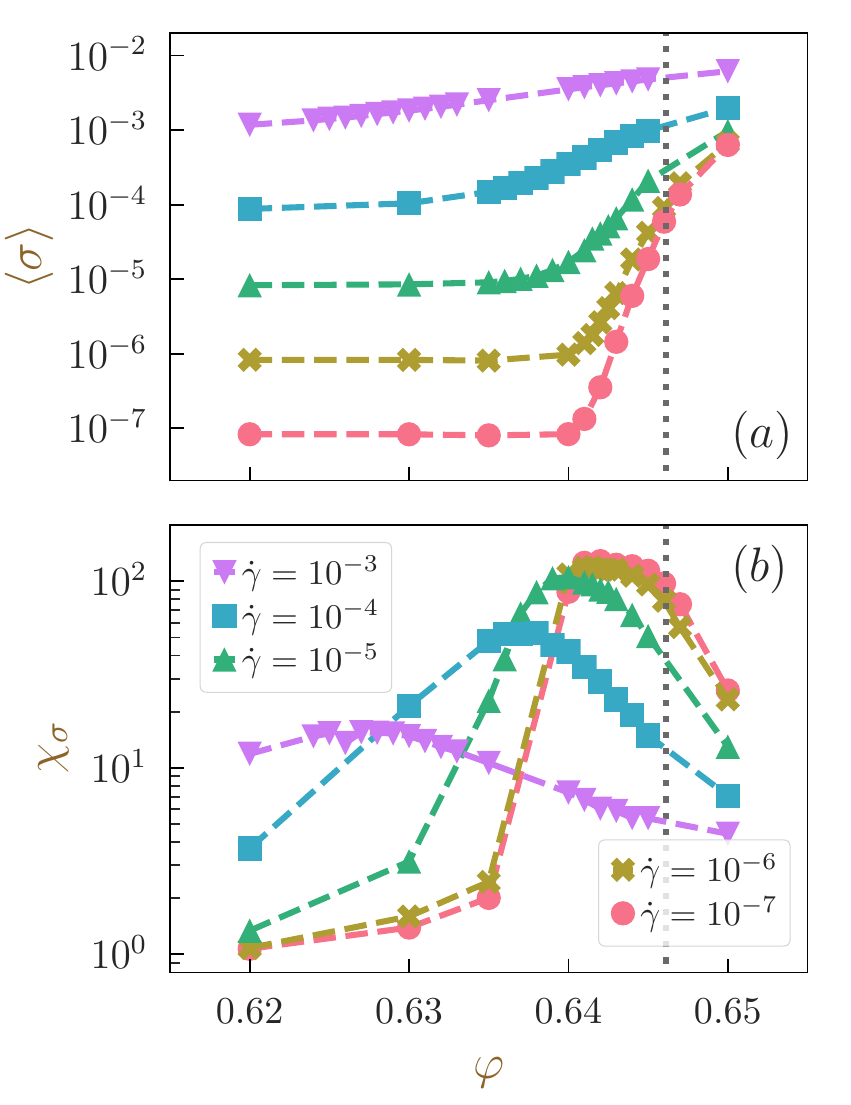}
\caption{\label{fig:stress_fluc}
Simulation results of (a) the average stress $\langle\sigma\rangle$ and (b) the susceptibility $\chi_\sigma$ as functions of the volume fraction $\varphi$.
Different markers are used to distinguish the different shear rate $\dot{\gamma}$ values, as shown in the legend in (b).
The dotted lines depict the location of the jamming point $\varphi_{\rm J}$.
}
\end{figure}

%------------------------------------------------------------------------------------------
\subsection{Average shear stress}\label{sec:stress_ave}
%------------------------------------------------------------------------------------------
We first present the dependence of the average shear stress $\langle \sigma\rangle$ on the volume fraction $\varphi$ and the shear rate $\dot{\gamma}$ in Fig.~\ref{fig:stress_fluc}a.
Specifically, the values of $\langle\sigma\rangle$ under different combinations of the parameters as functions of $\varphi$ are shown. 

% low varphi
In the low $\varphi$ regime, $\langle\sigma\rangle$ plateaus for all $\dot{\gamma}$.
We can also tell that $\sigma_{\rm low}$ scales linearly with $\dot{\gamma}$.
{This Newtonian-like shear rate dependence is considered the consequence of the effective overdamped dynamics due to the zero restitution coefficient.}

% high varphi
In the high $\varphi$ regime, $\langle \sigma\rangle$ increases with increasing $\varphi$.
In particular, when $\varphi$ is high enough and the system exhibits a clear yielding behavior, the $\dot{\gamma}$ dependence of $\langle\sigma\rangle$ follows the famous Herschel-Bulkley law~\cite{Herschel1926}:
$\langle\sigma\rangle\sim \sigma_{\rm Y}+\dot{\gamma}^n$ (see Appendix \ref{ap:HB} for the flow curve when $\varphi=0.65$~\cite{Kobayashi1980,Heussinger2009Phys.Rev.Lett.,Lin2014b,Saitoh2019a,Oyama2020a}).

% near the ``transition''
Between these two qualitatively different volume fraction regimes,
we observe a steep growth in $\langle\sigma\rangle$. 
As intuitively expected, this sharp increase in $\langle\sigma\rangle$ is observed in the vicinity of the jamming point ({$\varphi_{\rm J}\approx 0.6461$}; see Appendix \ref{ap:phij} for the determination of $\varphi_{\rm J}$ under shear~\cite{Kawasaki2015Phys.Rev.E,Kawasaki2020}).
However, 
the stress growth is most prominent at a volume fraction that is clearly smaller than $\varphi_{\rm J}$ {at finite $\dot{\gamma}$.}
Furthermore, as $\dot{\gamma}$ increases, the growth becomes less steep, and the onset volume fraction of the stress growth shifts towards the low $\varphi$ side.

%------------------------------------------------------------------------------------------
\subsection{Susceptibility of the shear stress}\label{sec:chi}
%------------------------------------------------------------------------------------------

We next focus on the fluctuation of the shear stress.
In particular, we quantify the enhancement of the collectivity in the fluctuations that accompanies the rapid increase in $\langle\sigma\rangle$ by the susceptibility $\chi_\sigma$, defined as:
\begin{align}
    \chi_{\sigma}\equiv N(\langle \sigma^2\rangle - \langle\sigma\rangle^2)/(\langle \sigma_{\rm local}^2\rangle-\langle\sigma_{\rm local}\rangle^2),  \label{eq:chi}  
\end{align}
{where $\langle \sigma_{\rm local}\rangle$ is the time- and particle-averaged value of the particle-based local stress $\sigma_i\equiv-\frac{N}{2V}\sum_{j\in{\rm contact}}f_{ijx}^{\rm el}(t)r_{ijy}(t)$ and $\sum_{j\in{\rm contact}}$ is the sum over the neighbors ($\langle\sigma_{\rm local}^2\rangle$ is the corresponding second-order moment)\footnote{
The only difference between the definitions of $\langle\sigma^2\rangle$ and $\langle\sigma_{\rm local}^2\rangle$ is the order in which the averages are taken over particles and time.}.}
With this definition, the average of $\sigma_i$ over the particles is identical to the macroscopic value $\sigma$, $\sigma=\frac{1}{N}\sum_i^N\sigma_i$.
This susceptibility $\chi_{\sigma}$ quantifies the degree of collectivity in the stress fluctuations: $\chi_\sigma$ is expected to diverge with increasing system size $N$ when the whole system behaves collectively, as in a system located near a critical point.
In Fig.~\ref{fig:stress_fluc}b, we plot the measurement results of $\chi_\sigma$ as a function of the volume fraction $\varphi$.

% low varphi
In the low $\varphi$ regime, $\chi_\sigma$ increases with increasing $\dot{\gamma}$.
However, interestingly, for low rates ($\dot{\gamma}\le 10^{-5}$), $\chi_\sigma$ hardly depends on $\dot{\gamma}$.
This behavior is in contrast to that of $\langle\sigma\rangle$, which depends linearly on $\dot{\gamma}$ for all shear rates $\dot{\gamma}$ in the low $\varphi$ regime. 
Regarding the volume fraction dependence in this regime,
$\chi_\sigma$ grows weakly with increasing volume fraction.

% high varphi
{In the high $\varphi$ regime, the opposite trend is observed: $\chi_\sigma$ becomes smaller when either $\dot{\gamma}$ or $\varphi$ increases.}
Still, the $\dot{\gamma}$ dependence disappears for low values of $\dot{\gamma}$ (in this case, $\dot{\gamma}\le 10^{-6}$), in accordance with the behavior in the low $\varphi$ regime.

% near the ``transition''
At an intermediate value of $\varphi$ between these two regimes, $\chi_\sigma$ exhibits a clear peak. 
As the shear rate $\dot{\gamma}$ increases, the {height of the peak decreases,}
%becomes shorter
and the position shifts towards the low $\varphi$ direction.
Note that if we further increase the shear rate to $\dot{\gamma}=10^{-2}$, we no longer observe a peak, at least in the range of the volume fraction that we have investigated, i.e., $0.62\le \varphi\le0.65$.
In accordance with the convergence of $\chi_\sigma$ both in the high and low $\varphi$ regimes, the height and position of the peak become almost constant for $\dot{\gamma}\le 10^{-6}$.
This total convergence of the susceptibility $\chi_\sigma$ in the low rate regime over all values of $\varphi$ suggests that the length scale that governs the stress fluctuation spans the whole system in this regime.
We discuss the possible candidates for this length scale in Sec.~\ref{sec:contact}, although we leave the precise identification for a future study.

Hereafter, we call the height and position of this peak $\chi_\sigma^{\rm max}(\dot{\gamma})$ and $\varphi_{\chi_\sigma^{\rm max}}(\dot{\gamma})$, respectively (we omit the explicit notation for the $\dot{\gamma}$ dependence below).

%------------------------------------------------------------------------------------------
%%% Fig. 2. Stress-strain curves (log)
%------------------------------------------------------------------------------------------
\begin{figure}
\includegraphics[width=\linewidth]{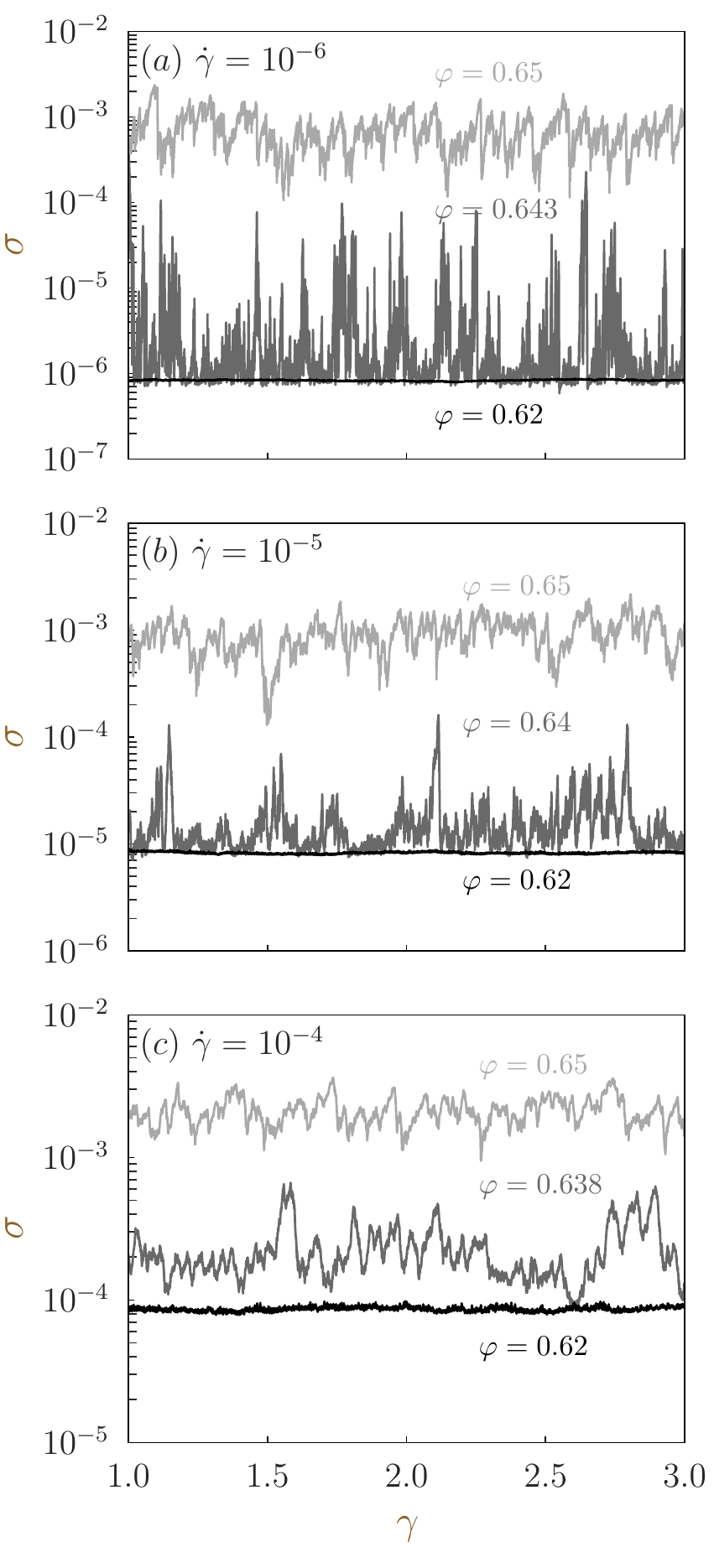}
\caption{\label{fig:s_s_log}
Stress-strain curves for 
{various combinations}
%systems with different combinations 
of the shear rate $\dot{\gamma}$ and the volume fraction $\varphi$.
The vertical axis follows a log-scale.
Results for (a) $\dot{\gamma}=10^{-6}$, (b) $\dot{\gamma}=10^{-5}$, and (c) $\dot{\gamma}=10^{-4}$ are shown.
In all panels, the results for three volume fractions are compared: $\varphi=0.62$ (black), $\varphi=0.65$ (light gray), and $\varphi=\varphi_{\chi_\sigma^{\rm max}}$ (dark gray).
}
\end{figure}

%------------------------------------------------------------------------------------------
\subsection{Stress-strain curves}\label{sec:s_s}
%------------------------------------------------------------------------------------------
To further obtain an intuitive understanding of the parameter dependence of the susceptibility $\chi_\sigma$, we plot typical stress-strain curves for the systems under {various}
%different 
combinations of the volume fraction $\varphi$ and the shear rate $\dot{\gamma}$ ($\dot{\gamma}=10^{-6},10^{-5}$ and $10^{-4}$) in Fig.~\ref{fig:s_s_log}.
For the whole parameter space investigated here, the average stress $\langle\sigma\rangle$ becomes larger
with both increasing $\varphi$ and increasing $\dot{\gamma}$, as presented in Fig.~\ref{fig:stress_fluc}a.
{However, the dependence on $\dot{\gamma}$ changes significantly depending on $\varphi$:
while the order of $\langle\sigma\rangle$ remains the same regardless of the value of $\dot{\gamma}$ at a high volume fraction ($\varphi=0.65>\varphi_{\chi_\sigma^{\rm max}}$, Fig.~\ref{fig:s_s_log} light gray curves)\footnote{Note that since $\langle\sigma\rangle$ obeys the Herschel-Bulkley law at this high volume fraction ($\varphi=0.65$), the order of the stress becomes larger if we apply a much faster shear.},}
it scales linearly with $\dot{\gamma}$ at a low volume fraction ($\varphi=0.62<\varphi_{\chi_\sigma^{\rm max}}$, Fig.~\ref{fig:s_s_log} black curves).
{However, the shapes of the stress-strain curves in these different regimes are similar in that the fluctuations are suppressed.}

{By contrast, the shape of the stress-strain curves dramatically changes in the vicinity of $\varphi_{\chi_\sigma^{\rm max}}$ under a slow shear rate ($\dot{\gamma}=10^{-6}$; Fig.~\ref{fig:s_s}a dark gray curve):}
we observe spiky peaks, with the height of the baseline being on the order of the stress at low $\varphi$ (see Appendix \ref{ap:s_s} for normal plots of the stress-strain curves where the spiky shapes are more appreciable).
The heights of the spikes are larger than the baseline by at most two orders of magnitude and barely reach the curve for $\varphi=0.65$.
Importantly, the probability distribution of $\sigma$, $P(\sigma)$, exhibits a power-law-like shape for $\varphi=\varphi_{\chi_\sigma^{\rm max}}$ and $\dot{\gamma}=10^{-6}, 10^{-5}$, indicating that this susceptibility peak reflects the criticality expected for $\dot{\gamma}\to 0$ {(see Appendix \ref{ap:s_s})}.
As the shear rate increases, the spikes become less sharp and less frequent ($\dot{\gamma}=10^{-5}$; Fig.~\ref{fig:s_s}b dark gray curve), and finally, the whole stress-strain curve becomes almost detached from that for a low $\varphi$ at $\dot{\gamma}=10^{-4}$ (Fig.~\ref{fig:s_s}c).
Since the magnitudes of the stress at the baseline and the peak top are comparable to those for low and high volume fractions respectively, we consider that these spikes are formed because the system goes back and forth between fluid-like low-stress states and solid-like large-stress states.
{That is, the whole system collectively changes its ``state" during the time evolution, as indicated by the susceptibility peak.}
We mention that similar repetitive transitions between fluid-like and solid-like states have also been observed under the AQS shear ($\dot{\gamma}=0$)~\cite{Heussinger2009Phys.Rev.Lett.}.
Notably, under a high shear rate ($\dot{\gamma}=10^{-4}$), $P(\sigma)$ exhibits a clear unimodal shape without power-law tails at either end {(see Appendix \ref{ap:s_s})}.
This observation suggests that the increase in $\sigma$ becomes more similar to a cross-over rather than a phase transition because of the effect of the strong external field (see Sec.~\ref{sec:Ising} for the qualitative similarity between our system and the conventional critical phenomena).

%%%%%%%%%%%%%%%%%%%%%%%%%%%%%%%%%%%%%%%%%%%%%%%%%%%%%%%%%%%%%%%%%%%%%%%%%%%%%%%%%%%%%%%%%%%%%%%%%%%%%%%%%%%%%%%%%%%%%%%%%%%%%
% Discussion
%%%%%%%%%%%%%%%%%%%%%%%%%%%%%%%%%%%%%%%%%%%%%%%%%%%%%%%%%%%%%%%%%%%%%%%%%%%%%%%%%%%%%%%%%%%%%%%%%%%%%%%%%%%%%%%%%%%%%%%%%%%%%%
\section{Discussion}
In this section, we discuss the similarity between our system and the conventional critical phenomenon: the ferromagnetic transition in the Ising model under an external field.
Based on this analogy, we can tell that the shear stress $\sigma$ can be viewed as a natural ``conjugate" variable to the strength of the external field (namely, the shear rate $\dot{\gamma}$).
However, $\sigma$ changes its value by orders of magnitude depending on $\dot{\gamma}$ even in the ``disordered", low-stress phase.
In this sense, it is qualitatively different from conventional standard order parameters that are normalized to be between zero and one in most cases.
Therefore, we further conduct the same analysis for an alternative candidate for an order parameter, i.e., the interparticle contact number $z$.

%------------------------------------------------------------------------------------------
\subsection{Correspondence to conventional criticality in equilibrium systems}\label{sec:Ising}
%------------------------------------------------------------------------------------------

To further explore the parameter dependence of the shear stress $\sigma$ and its fluctuations, we rely on an analogy with a well-understood phase transition.
Here, in particular, we discuss an analogy with one of the most famous examples: the Ising model under a magnetic field {(see Appendix \ref{ap:ising} for a brief recapitulation of the mean-field solution)}.
As shown in Figs.~\ref{fig:stress_fluc}(a,b), the average and the susceptibility of the stress exhibit qualitative similarities with the magnetization and the susceptibility in the Ising model (Appendix \ref{ap:ising}):
{the inverse temperature $\beta$, which is the control parameter of the criticality in the Ising model, corresponds to the volume fraction $\varphi$ in our system. Similarly, the external magnetic field $h$ and the magnetization $m$ correspond to the shear rate $\dot {\gamma}$ and the mean stress $\langle \sigma \rangle$, respectively.}
Moreover, in both systems, as the external field ($h$ or $\dot{\gamma}$) increases, the change in the order parameter ($m$ or $\langle\sigma\rangle$) becomes less steep, and the whole plot shifts towards the less-ordered side.
Regarding the susceptibility ($\chi$ or $\chi_\sigma$), we observe peaks at a value of the control parameter ($\beta$ or $\varphi$) that is shifted from the critical point when an external field is present. The height of these peaks decreases with increasing external field, and the position shifts towards the small-order side.
{We emphasize that the counterpart of the magnetic field in our system is not the strain $\gamma$ but the shear rate $\dot {\gamma}$, which is the conjugate of the stress in effective energy dissipation. 
Hence, the free energy of the Ising model corresponds to the dissipation function in our system and is consistent with the empirical knowledge that the dissipation system takes precedence over the dynamics of the minimum energy dissipation~\cite{Unger2004,Torok2007}.}
In this sense, the shear stress $\sigma$ can be viewed as a natural conjugate variable to the external field and thus as an order parameter.
However, since $\sigma$ is dependent not only on the existence of contacts but also on the degree of overlapping of each contact, it changes its value by orders of magnitude depending on $\dot{\gamma}$ even in the ``disordered", dilute state.
In the next section, we instead measure the average and susceptibility of the interparticle contact number, the values of which are expected to exhibit less $\dot{\gamma}$ dependence.

%------------------------------------------------------------------------------------------
\subsection{Contact number}\label{sec:contact}
\begin{figure}
\includegraphics[width=\linewidth]{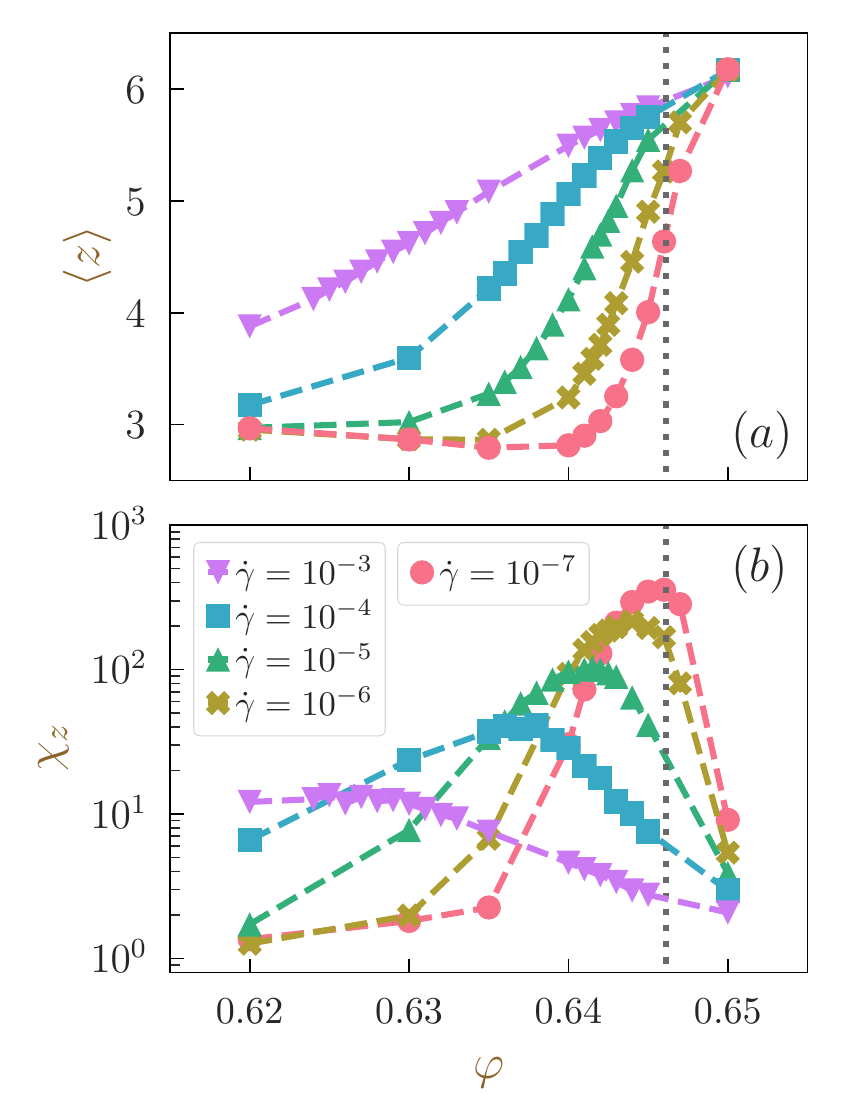}
\caption{\label{fig:contact_nematic}
{
  (a) Average $\langle z\rangle$ and
  (b) susceptibility $\chi_z$ of the interparticle contact number as functions of the volume fraction
  $\varphi$.
  Different markers represent different shear rates, as shown in the
  legend in Fig.~\ref{fig:stress_fluc}(b).
  The dotted lines depict the location of the jamming point $\varphi_{\rm J}$.
}
}
\end{figure}

The interparticle contact number $z$ characterizes the jamming transition most directly in terms of the microscopic structures~{\cite{VanHecke2010}}.
For the jamming transition in quiescent systems without external fields, $z$ changes discontinuously from zero to approximately $z_{\rm C}$ at the critical point $\varphi_{\rm J}$, above which physical quantities such as the pressure or the shear modulus change in a power-law manner, as in the case of the conventional second-order phase transitions~{\cite{OHern2003}}.
According to Maxwell's condition, $z_{\rm C}=2d$ holds for frictionless soft athermal spheres, where $d$ is the spatial dimension of the system.
Here, we plot the average and the susceptibility of the interparticle contact number $z$ (we do not exclude rattlers to compute $z$) under a finite-rate shear as functions of $\varphi$ in Fig.~\ref{fig:contact_nematic}.
For the definition of the susceptibility $\chi_z$, we employ a definition similar to Eq.~\ref{eq:chi}.

% low shear rate
The dependence of the average contact number $\langle z\rangle$ on the volume fraction $\varphi$ is qualitatively very similar to that of the average stress $\langle\sigma\rangle$: it is almost constant in the low $\varphi$ regime and then shows sudden growth around $\varphi_{\rm J}$, after which the growth rate decreases in the high $\varphi$ regime.
However, the dependence on $\dot{\gamma}$ is significantly different from that of $\langle\sigma\rangle$: in the low $\varphi$ regime, the plateau disappears for high $\dot{\gamma}$, and the shear rate dependence is not linear.
Furthermore, the values of $\langle z\rangle$ at the highest $\varphi$ hardly depend on $\dot{\gamma}$.

The susceptibility of the contact number $\chi_z$ behaves qualitatively very similarly to that of $\chi_\sigma$: it exhibits a clear peak near the jamming point $\varphi_{\rm J}$, and the peak height and position change in the same way as $\chi_\sigma$ when $\dot{\gamma}$ increases.
One major difference from $\chi_\sigma$ is that the peak position and height of $\chi_z$ obviously change even in the low rate limit $\dot{\gamma}\le 10^{-6}$, where $\chi_\sigma$ becomes constant.

{This qualitative difference intriguingly suggests that the characteristic lengths that govern $\sigma$ and $z$ ($\xi_\sigma$ and $\xi_z$, respectively) are different.
Let us enumerate several candidates from previous studies.
For example, it is known that the correlation length of the deviation from the continuum description diverges at the jamming point~\cite{Ellenbroek2006,Ellenbroek2009,Lerner2014,Mizuno2016}.
This length scale, often referred to as $l_c$, is a candidate for $\xi_z$.
On the other hand, the isotropic as-quenched state has recently been shown to be qualitatively different from the sheared nonequilibrium steady state in terms of the stability against perturbation, even in the AQS limit ($\dot{\gamma}=0$)~\cite{Karmakar2010a,Oyama2020}.
This knowledge implies that $l_c$ and $\xi_z$ can be different in nature, since $l_c$ is measured in the absence of an external field ($\gamma=0$), while $\xi_z$ should be measured in the steady state $\gamma>{\cal O}(1)$.
As an example of a correlation length measured in a dynamic situation, Refs.~\cite{Saitoh2016b,Saitoh2016c,Saitoh2020} reported that the correlation length of the nonaffine velocities of particles diverges in the limit of $\varphi\to\varphi_{\rm J}$ and $\dot{\gamma}\to 0$ in two-dimensional packings of soft frictionless disks.
However, this correlation length has been shown to remain finite even in the same limit in three dimensions~\cite{Oyama2019}.
Instead, in ref.~\cite{Oyama2019}, the authors introduced the correlation length of the vortex clusters, which diverges in that limit.
As another example of a dynamical correlation length, the one associated to the yielding criticality is also known to diverge in the limit of $\dot{\gamma}\to 0$~\cite{Lin2014b,Oyama2020a}.
However, this length scale can be well defined only in the high $\varphi$ regime, where the Herschel-Bulkley law is valid and cannot describe the total convergence of $\chi_\sigma$ over the whole $\varphi$ regime.
As discussed here, multiple candidates exist, with the possibility that none of them is the desired one.
Although identifying the governing length scale by comparing all these candidates is an important issue, we leave it as a future problem.}

\begin{figure}
\includegraphics[width=\linewidth]{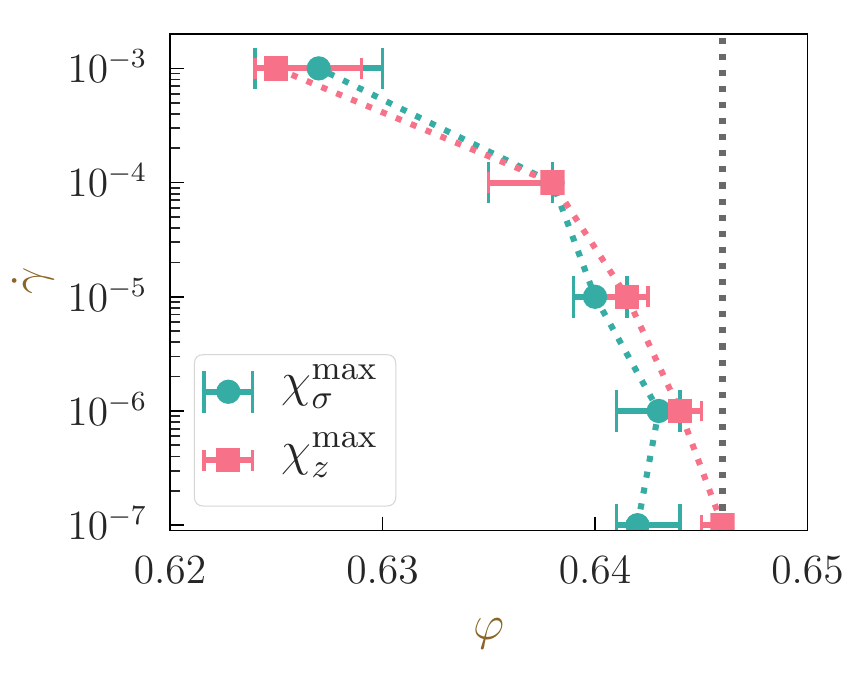}
\caption{\label{fig:phase_diagram}
{Widom lines, or a plot of the locations of the susceptibility peaks $\varphi_{\chi_\alpha^{\rm max}}$ on a $\varphi-\dot{\gamma}$ plane, where $\alpha\in\{\sigma, z\}$. 
Different symbols represent different definitions of the susceptibility, as shown in the legend.
The dotted line shows the location of the jamming point $\varphi_{\rm J}\approx 0.6461$ estimated at $\dot{\gamma}=0$ (see Appendix \ref{ap:phij}).
Error bars indicate the range of $\varphi$ for which the values of $\chi_\alpha$ are greater than 90\% of $\chi_\alpha^{\rm max}$.}
}
\end{figure}

Finally, we present the ridges obtained by connecting the peaks of the susceptibilities under different values of $\dot{\gamma}$ in Fig.~\ref{fig:phase_diagram}.
In this plot, we compare the results for $\chi_\sigma$ and $\chi_z$.
These ridges can be regarded as the dissipative-system counterpart of the Widom lines by definition.
Both Widom lines seem to converge to $\varphi_{\rm J}$ in the limit of $\dot{\gamma}\to 0$, as expected.
Moreover, these two lines follow different paths, as is the case for the conventional equilibrium systems, e.g., the Widom lines around the liquid-gas critical point.

%------------------------------------------------------------------------------------------
%%% Fig. ４. Contact Number & Nematic Order
%------------------------------------------------------------------------------------------

%------------------------------------------------------------------------------------------
%%% Fig. 3. Phase Diagram
%------------------------------------------------------------------------------------------
%------------------------------------------------------------------------------------------
%------------------------------------------------------------------------------------------

%%%%%%%%%%%%%%%%%%%%%%%%%%%%%%%%%%%%%%%%%%%%%%%%%%%%%%%%%%%%%%%%%%%%%%%%%%%%%%%%%%%%%%%%%%%%%%%%%%%%%%%%%%%%%%%%%%%%%%%%%%%%%%
% Summary
%%%%%%%%%%%%%%%%%%%%%%%%%%%%%%%%%%%%%%%%%%%%%%%%%%%%%%%%%%%%%%%%%%%%%%%%%%%%%%%%%%%%%%%%%%%%%%%%%%%%%%%%%%%%%%%%%%%%%%%%%%%%%%
\section{Summary and overview}
In this work, we conducted MD simulations for dense packings of soft athermal spheres under a finite-rate shear and investigated the dependence of the statistics of the shear stress on the shear rate and the volume fraction.
The average stress changes largely in the vicinity of the jamming point; moreover, the onset volume fraction for the stress growth becomes smaller when the shear rate increases.
{Interestingly, this sudden stress growth is accompanied by the formation of a peak of the susceptibility.}
{To further understand this susceptibility peak, we investigated the time evolution of the stress. We found that the stress-strain curve exhibits spiky peaks at the volume fraction where the susceptibility peak is observed. These peaks are formed since the system can temporally gain solidity with the aid of the external shear, while it is fluidic otherwise. }
We furthermore measured the average and susceptibility of the interparticle contact number as an example of a normalized order parameter in our system.
The results for $\chi_z$ are qualitatively consistent with those for $\chi_\sigma$, although the length scales that govern these two fluctuations seem different.
We furthermore visualized the Widom lines in our system, or the ridges of the susceptibility peaks for both the stress and contact number.
As the equilibrium phase diagram shows, two Widom lines follow different paths, although both seem converge to a critical point in the limit $\dot{\gamma}\to 0$.

As a future direction, an investigation of whether modification of the physical dimension~\cite{Radjai2002,Saitoh2016c,Oyama2019}, the damping coefficient~\cite{Andreotti2012,Kawasaki2014a,Vagberg2017}, or the local dissipation mechanisms (e.g., introduction of the tangential friction~\cite{Otsuki2009a,Otsuki2011}) leads to any qualitative changes should be carried out.

\begin{acknowledgments}
  We thank Atsushi Ikeda, Kota Mitsumoto, and Yusuke Hara for the fruitful discussions.
This work was financially supported by JSPS KAKENHI Grant Numbers 18H01188, 18K13464, 19K03767, 20H05157, 20H00128, 20H01868, 20J00802, and 20K14436.
\end{acknowledgments}

\begin{appendix}

\section{Flow curve for a dense system}\label{ap:HB}
\begin{figure}
\includegraphics[width=\linewidth]{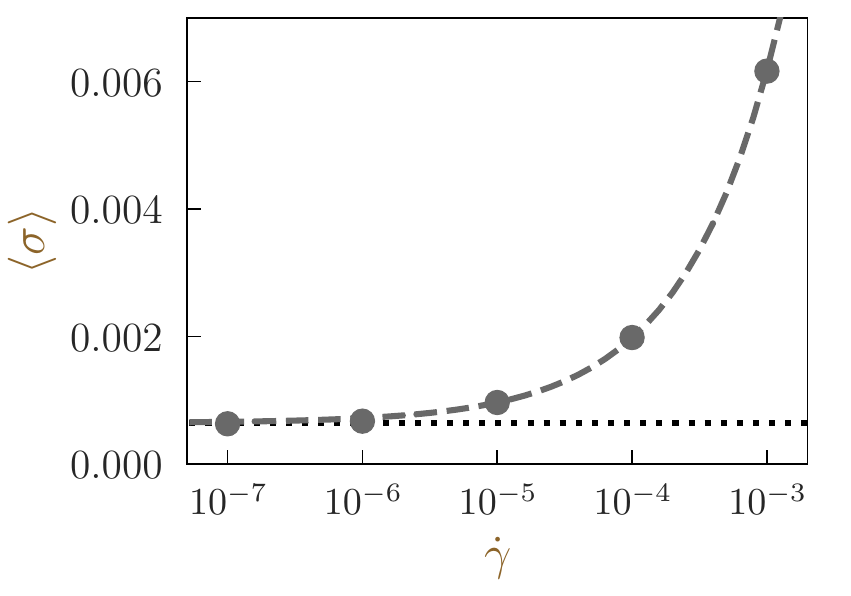}
\caption{\label{fig:flow_curve}
The dependence of the average stress $\langle\sigma\rangle$ on the shear rate $\dot{\gamma}$ at $\varphi=0.65$.
The markers represent the results of the MD simulation, and the dashed line is the fitting to the Herschel-Bulkley law. The Herschel-Bulkley exponent here is $n\sim 0.62$.
The dotted line depicts the value of $\langle\sigma\rangle$ measured under the AQS shear.
}
\end{figure}

In Fig.~\ref{fig:flow_curve}, we plot the average shear stress $\langle\sigma\rangle$ in the system as a function of the shear rate $\dot{\gamma}$ with $\varphi=0.65$.

To roughly estimate the yield stress $\sigma_{\rm Y}$, we also conduct an AQS simulation.
In the AQS simulation, instead of integrating the equation of motion, we minimize the potential energy of the system~\cite{Kobayashi1980,Heussinger2009Phys.Rev.Lett.,Saitoh2019a}.
We employ the FIRE algorithm~\cite{Bitzek2006} and terminate the iteration when the maximum magnitude of the force exerted on one particle meets $f_{\rm max}<10^{-9}$.
The strain increment is $\Delta\gamma=5\times 10^{-5}$~\cite{Heussinger2009Phys.Rev.Lett.}.

Fig.~\ref{fig:flow_curve} shows that the average stress converges to the AQS value at very slow shear rates ($\dot{\gamma}\le 10^{-6}$).
In other words, these shear rates can be considered as in the so-called \emph{quasistatic regime}.
We regard the average stress under the AQS shear as the yield stress and further fit the numerical results to the Herschel-Bulkley law, $\langle\sigma\rangle\sim \sigma_{\rm Y}+\dot{\gamma}^n$.
This simple estimation provides $n\sim 0.62$, and the obtained curve captures the numerical data very well.
Note, however, that we must take into account the finite size effects to accurately evaluate the Herschel-Bulkley parameters, namely, the yield stress and the critical exponent~\cite{Lin2014b,Oyama2020a}.

%------------------------------------------------------------------------------------------
%%% Figure S2: Flow curve for the system with $\varphi=0.65$
%------------------------------------------------------------------------------------------
%------------------------------------------------------------------------------------------
%------------------------------------------------------------------------------------------

\section{Determination of the jamming point}\label{ap:phij}
\begin{figure}
\includegraphics[width=\linewidth]{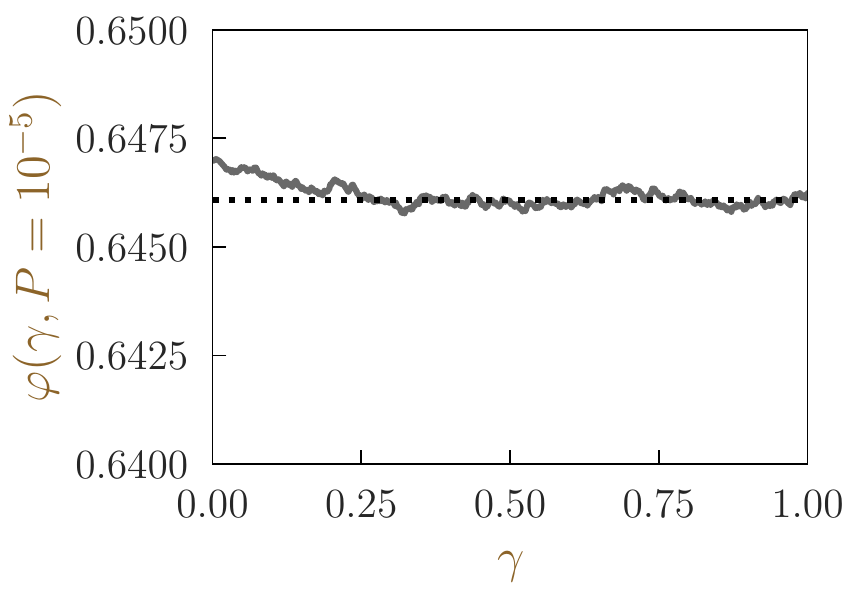}
\caption{\label{fig:jamming_point}
Average over 60 samples.
Dashed line represents the estimated value of $\varphi_{\rm J}=0.6461$ (see the text regarding how to locate it).
}
\end{figure}

We determine the precise location of the jamming point $\varphi_{\rm J}$ following ref.~\cite{Kawasaki2020}, the protocol of which we briefly explain here.
We first prepare a random particle configuration in a fixed volume system with a linear dimension $L$.
Then, we minimize the total potential energy of the system to obtain a mechanically equilibrated configuration, with the pressure $p$ being controlled such that $p\approx 0$.
We furthermore apply shear in an AQS manner until the system reaches a steady state ($\gamma=1$) with a strain increment $\Delta\gamma=10^{-3}$.
The volume fraction at the steady state ($\gamma>0.5$) can be used as a well-defined jamming point~\cite{Kawasaki2020}.
{With this protocol, the value of the jamming point does not depend on the initial configuration.}
Because achieving numerically the exact mechanically equilibrated configuration with zero pressure is almost impossible, we set the target pressure $P=10^{-5}$.
Following this protocol and averaging over the values in the steady state ($\gamma\ge 0.5$), we locate the jamming point as {$\varphi_{\rm J}\approx 0.6461$ (Fig.~\ref{fig:jamming_point})}.
This value of $\varphi_{\rm J}$ is consistent with the one estimated by directly fitting the diverging trend of the viscosity~\cite{Kawasaki2015Phys.Rev.E}.
We stress that the data in Fig.~\ref{fig:jamming_point} is the average over 60 samples.

%------------------------------------------------------------------------------------------
%%% Figure S2: Flow curve for the system with $\varphi=0.65$
%------------------------------------------------------------------------------------------
%------------------------------------------------------------------------------------------
%------------------------------------------------------------------------------------------

\section{Stress-strain curves in a normal plot}\label{ap:s_s}
\begin{figure*}
\includegraphics[width=\linewidth]{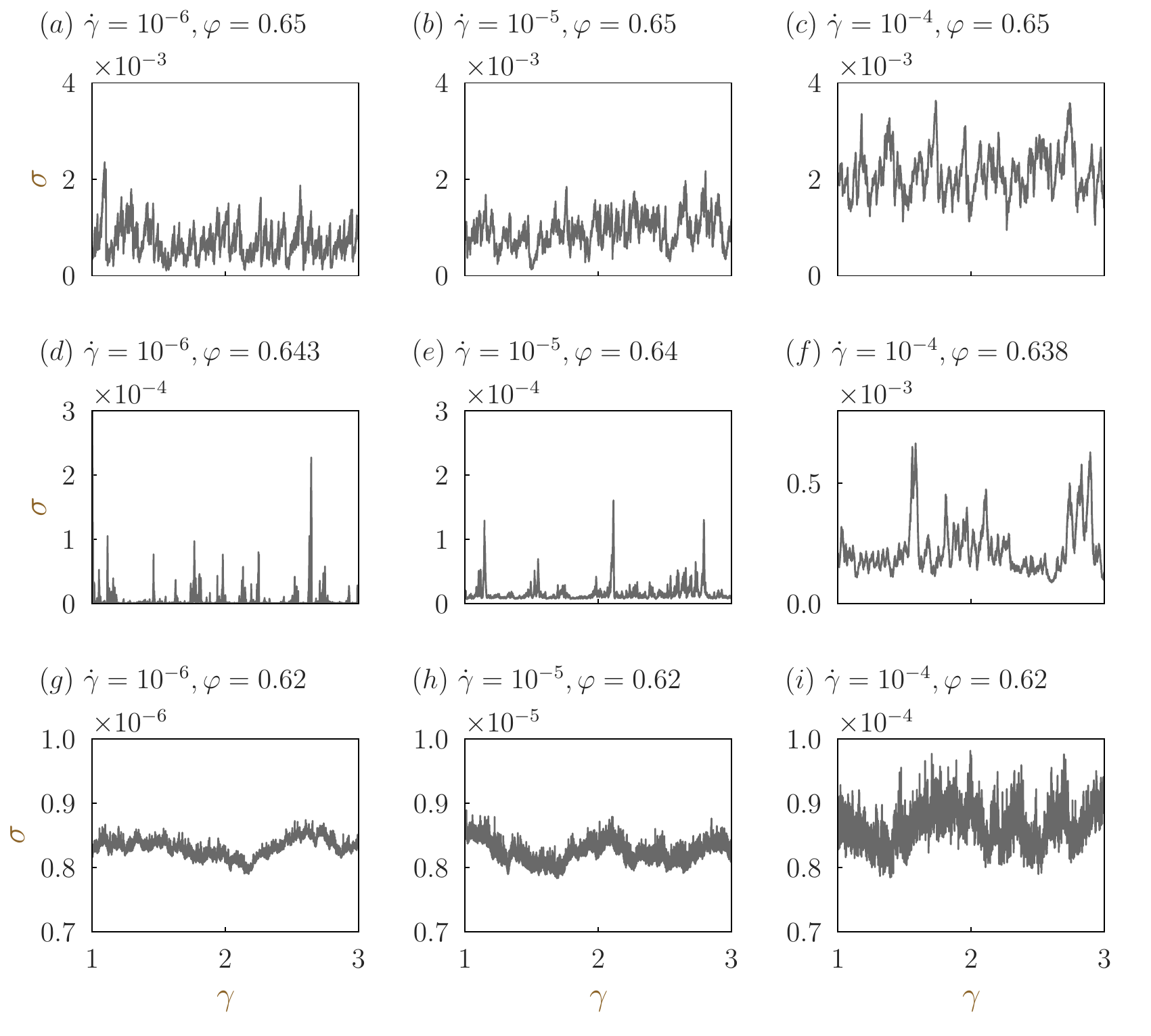}
\caption{\label{fig:s_s}
Stress-strain curves for systems with different combinations of the shear rate $\dot{\gamma}$ and the volume fraction $\varphi$.
{Results for (top row) $\varphi=0.65>\varphi_{\chi_\sigma^{\rm max}}$}, (middle row) $\varphi=\varphi_{\chi_\sigma^{\rm max}}$, and (bottom row) $\varphi=0.62<\varphi_{\chi_\sigma^{\rm max}}$ are shown.
From left to right, the shear rate increases as follows: $\dot{\gamma}=10^{-6}, 10^{-5}$ and $10^{-4}$.
}
\end{figure*}

In this section, we present a normal plot of the stress-strain curves for various combinations of parameters in Fig.~\ref{fig:s_s} (the ones used for Fig.~2 in the main text are employed).
In the plots for $\varphi=\varphi_{\chi_\sigma^{\rm max}}$, we observe sharp spikes, especially for a slow shear rate.

\begin{figure}
\includegraphics[width=\linewidth]{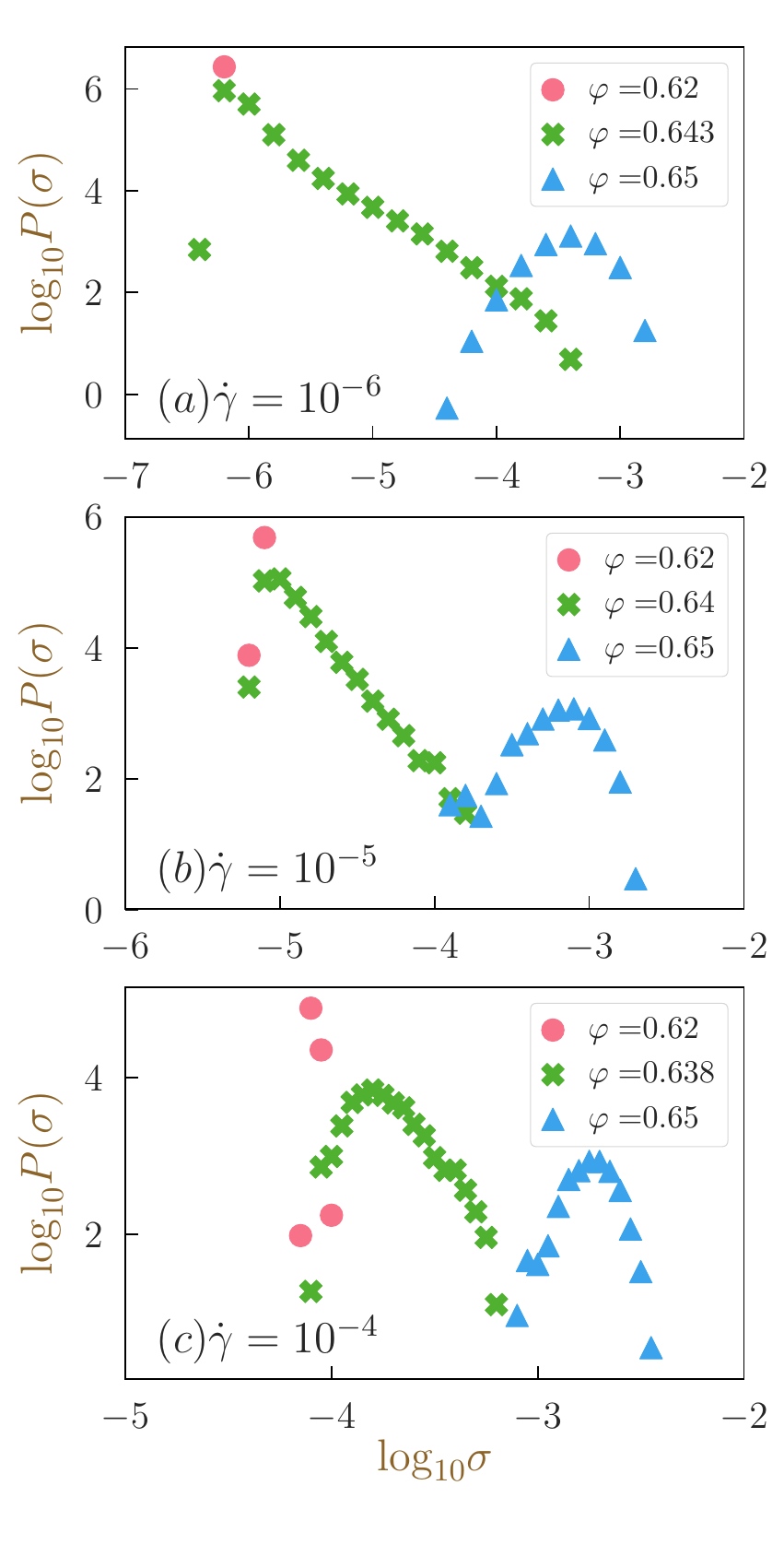}
\caption{\label{fig:pdf_s}
PDFs of the shear stress $\sigma$ for different shear rates $\dot{\gamma}=10^{-6}, 10^{-5}$ and $10^{-4}$.
Different symbols indicate different volume fractions $\varphi$, as shown in the legend.
}
\end{figure}

In Fig.~\ref{fig:pdf_s}, we plot the probability distribution function (PDF) of the shear stress $\sigma$ for the same combinations of parameters.
At a low volume fraction $\varphi=0.62$, the PDF is almost Dirac's delta function for all shear rates (the width is very narrow).
At a high volume fraction $\varphi=0.65$, the PDF is unimodal, with a large width for all shear rates.
At $\varphi=\varphi_{\chi_\sigma^{\rm max}}$, however, we observe shear rate dependence.
Although the PDF exhibits a power-law-like shape for slow shear rates ($\dot{\gamma}\le 10^{-5}$), it becomes rather regular unimodal shape for a high shear rate ($\dot{\gamma}=10^{-4}$).

%------------------------------------------------------------------------------------------
%%% Fig. S3. Normal plots of stress-strain curves 
%------------------------------------------------------------------------------------------

%------------------------------------------------------------------------------------------
%%% Fig. S4. PDF of stress
%------------------------------------------------------------------------------------------

%------------------------------------------------------------------------------------------

\section{Ising model}\label{ap:ising}
\begin{figure}
\includegraphics[width=\linewidth]{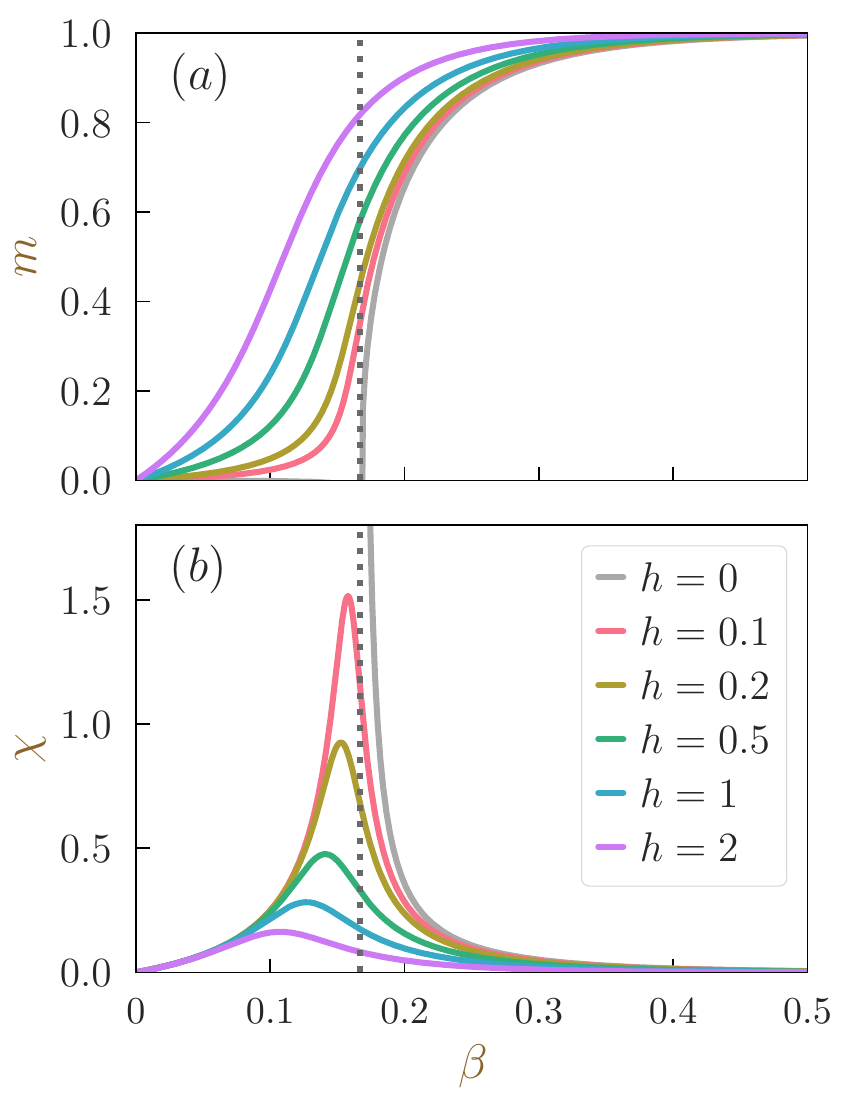}
\caption{\label{fig:ising_solution}
(a) Magnetization $h$ and (b) susceptibility $\chi$ obtained from the self-consistent Eq.~\ref{eq:SC} as functions of the inverse temperature $\beta$. Different colors represent different values of $h$, as shown in the legend.
}
\end{figure}

In this section, we recapitulate the famous self-consistent equation for the magnetization of the Ising model under an external magnetic field, which is derived with a mean-field approximation.
Assume we have a $d$-dimensional Ising-type spin system on a regular lattice whose Hamiltonian ${\cal H}$ is written as:
\begin{align}
    {\cal H}=-J\sum_{\langle ij\rangle}S_iS_j-h\sum_{i=1}^N S_i,
\end{align}
where $S_i\in\{1,-1\}$ is the spin variable at the site $i$, $h$ stands for the strength of the external field, and $J$ represents the coupling constant.
Then, with a mean-field approximation, we can derive a self-consistent equation for the spin $\frac{1}{2}$ magnetization $m$ as:
\begin{align}
    m=\text{tanh}(\beta Jz_cm+\beta h),\label{eq:SC}
\end{align}
where $z_c=2d$ is the spin coordination number.
If we employ $d=3$ and $J=1$, the critical inverse temperature $\beta_c$ is obtained as $\beta_c=1/Jz \approx 0.167$.
We plot the values of the magnetization $m$ and the susceptibility $\chi=dm/dh$ as functions of $\beta$ for various values of $h$ in Fig.~\ref{fig:ising_solution}.

%------------------------------------------------------------------------------------------
%%% Fig. S5. Ising solution
%------------------------------------------------------------------------------------------

\end{appendix}
	
%
%\bibliography{Sheared_Jamming}

%
%
\end{document}